\documentstyle[multicol,aps,prl,epsf]{revtex}

\begin{document}

\draft

\title{Electric field induced charge injection or exhaustion in organic 
thin film transistor}

\author{
Manabu Kiguchi$^1$, Manabu Nakayama$^1$, Toshihiro Shimada$^2$, Koichiro Saiki$^{1,2}$}

\address{$^1$Department of Complexity Science $\&$ Engineering, Graduate
School of Frontier Sciences, The University of Tokyo, 7-3-1 Hongo,
Bunkyo-ku,Tokyo 113-0033, Japan}

\address{$^2$Department of Chemistry, Graduate School of Science, The 
University of Tokyo, 7-3-1 Hongo, Bunkyo-ku, Tokyo 113-0033, Japan}

\date{\today}

\maketitle

\begin{abstract}
The conductivity of organic semiconductors is measured 
{\it in-situ} and continuously with a 
bottom contact configuration, as a function of film thickness at various gate voltages. 
The depletion layer thickness can be directly determined 
as a shift of the threshold thickness at which electric 
current began to flow. The {\it in-situ} and continuous measurement can 
also determine qualitatively 
the accumulation layer thickness together with the distribution 
function of injected carriers. 
The accumulation layer thickness is a few mono layers, 
and it does not depend on gate voltages, rather depends on the chemical species. 

\end{abstract}

\medskip


\begin{multicols}{2}
\narrowtext

\section{INTRODUCTION}
\label{sec1}

Charge injection or depletion by electric fields in semiconductors is 
the basis of field electric transistors (FET)\cite{1}. 
It is also the basis of the field effect control of physical properties, 
which is one of the hottest topics in the field of condensed matter science. 
In contrast with chemical doping, 
the charge injection by electric fields does not introduce 
any chemical or structural disorder in pristine materials\cite{2,3}, 
and it may be possible to inject carriers in host materials, to which charges 
can not be injected by chemical doping. In this context, determination of the region, 
where carriers are practically injected or depleted by electric fields, 
is an important and fundamental issue. 
However, there had been little investigation on evaluating experimentally 
the dimension of these 
regions\cite{4}.
 
Recently, Dodabalapur {\it et al.} reported a pioneering work 
relating to the accumulation layer 
thickness. They estimated that it was 1$\sim$2 ML by comparing FET characteristics of a 2 ML 
sexithiophene ($\alpha$-6T) film with that of bulk one\cite{5}. 
However, they did not quantitatively determine 
the thickness, and quantitative analysis has not been done yet. 
Recently, we proposed a simple 
method to determine quantitatively the thickness of the depletion and accumulation layers in 
organic semiconductor pentacene\cite{6}.
 The conductivity of pentacene thin film transistor (TFT) was 
measured {\it in-situ} and continuously with a bottom contact configuration, 
as a function of film thickness. 
However, we only discussed the thickness of these layers for one organic species at fixed 
gate voltage (0, $\pm$15 V). 
By comparing the result of pentacene TFT with that of another organic TFT, 
and that at various gate voltages, 
we could discuss the characteristics of the conductance in organic 
TFT in more detail. Furthermore, there are many points to discuss; 
way and validity of estimation, universality of the method, etc.
  
In this study, we have studied the {\it in-situ} 
and continuous measurement for pentacene and 6T. 
By measuring the gate voltage and chemical species dependence, 
we quantitatively discussed the 
thicknesses of the depletion and accumulation layers in detail, and characteristics of the 
conductance in organic materials. 

\section{EXPERIMENTAL}
\label{sec2}

\begin{figure}
\begin{center}
\leavevmode\epsfysize=40mm \epsfbox{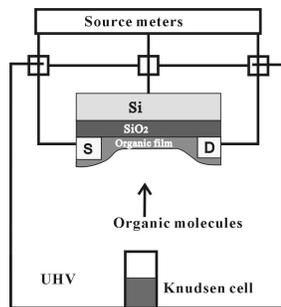}
\caption{Schematic sample layout. The source (S)-drain (D) current was 
measured {\it in-situ}  at various gate voltages.}
\label{fig1}
\end{center}
\end{figure}

Figure~\ref{fig1} shows the schematic sample layout. 
The substrate was a highly doped silicon wafer, acting as a gate electrode. 
The gate dielectric layer was a 700 nm thermally grown silicon dioxide. 
On top of the surface, 30 nm thick gold source(S) and drain(D) electrodes 
were deposited through a shadow mask. 
The channel length and width were 100 $\mu$m and 5.4 mm. 
Pentacene (Aldrich) and $\alpha$-6T (Syncom BV) were deposited 
by means of vacuum deposition.
The film thickness was monitored by a quartz crystal oscillator, 
and it was calibrated by Auger Electron Spectroscopy.
The deposition rate was 0.1 nm/min, 
and the substrate temperature was kept at 310 K during the growth. 
Under high-vacuum condition, the S-D current of TFT was measured 
as a function of film thickness with a bottom-contact configuration. 
During the growth of the organic films, electric fields were not 
applied in order to exclude the influence of the electric current on the film growth. 
As a function of the film thickness, 
the gradual change in the conductivity and FET characteristics were measured 
{\it in-situ} under the same condition, avoiding the problem of specimen dependence.

\section{RESULTS}
\label{sec3}
\subsection{Depletion layer}
\label{sec3a}
\begin{figure}
\begin{center}
\leavevmode\epsfysize=50mm \epsfbox{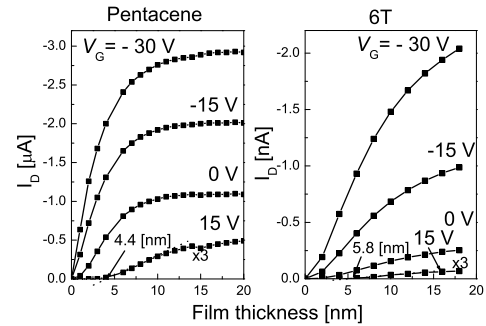}
\caption{Source-drain current ($I_{SD}$) of pentacene and 6T measured 
as a function of film thickness at various gate voltages ($V_{G}$). 
Source-drain voltage was kept at 1 V. The dotted line is guideline to 
obtain the threshold thickness.}
\label{fig2}
\end{center}
\end{figure}

 Figure~\ref{fig2} shows the S-D current ($I_{SD}$) of pentacene 
and 6T as a function of film thickness at various 
gate voltages ($V_{G}$); $V_{G}$ =-30, -15, 0, and +15 V. 
The S-D voltage ($V_{SD}$) was kept at 1 V. Since both 
pentacene and 6T are p-type semiconductors, 
large $I_{SD}$ is observed at negative gate voltages. Here 
we find the presence of threshold thickness ($d_{th}$) 
at which electric current begins to flow. 
There are two points to note about $d_{th}$. First, $d_{th}$ is $\sim$1 nm at $V_{G}$=0 V, 
and finite electric current is observed even for the 1.0 nm thick film. 
The present result indicates that the S, D electrodes (width 100 $\mu$m) 
could be electrically connected by only 1 nm thick organic semiconductors. 
Furthermore, typical p-type FET characteristics were observed for the this film \cite{7}. 
Second, $d_{th}$ shows clear $V_{G}$ dependence, 
and it shifts up to 4.4($\pm$0.5) nm for pentacene and $d_{th}$=5.8($\pm$0.6) nm 
for 6T at $V_{G}$=15 V. The value of $d_{th}$ was well reproduced for several experiments.
When positive $V_{G}$ is applied to p-type semiconductors, 
the depletion layer with a low conductance is 
formed at the semiconductor/insulator interface. 
Then, the whole film can be depleted on the thin 
limit, and $I_{SD}$ does not flow at positive $V_{G}$. 
Consequently, it is suggested that $d_{th}$ corresponds to the depletion layer thickness. 

In order to discuss the validity of the above discussion, 
we estimated the concentration of major carriers by two different ways 
with the conventional model which has been already established for 
inorganic semiconductors\cite{1,8}. 
When a depletion layer is formed at the semiconductor/insulator interface, 
carriers are exhausted and ionized acceptors are left in the depletion layer. 
The induced charge amounts to be $Q=CV_{G}$, 
if a flat band condition is satisfied at $V_{G}$=0 V. 
Here, $C$ is the capacitance of the gate dielectric, 
and it is represented as $C=\epsilon _{OX} / t_{OX}$, 
where $\epsilon _{OX}$: dielectric constant of 
SiO$_{2}$, $t_{OX}$: thickness of SiO$_{2}$. 
In the depletion approximation, only ionized acceptors are left in the 
depletion layer and distributed uniformly in the depletion layer. 
The concentration of the acceptors ($N$) (i.e. exhausted major carriers) is, 
thus, obtained by dividing induced charge ($Q$) by the depletion 
layer thickness ($t_{s}$), and it is represented as $N=Q/et_{s}$ . 
Because of the charge neutrality at $V_{G}$=0 V, 
concentration of the major carriers is equal to that of acceptors. 
Therefore, the concentration of the 
major carriers of pentacene at $V_{G}$=0 V amounted to $N$=$1.1 \times 10 ^{18} cm^{-3}$ 
using $t_{s}$=4.4 nm at $V_{G}$=15 V, 
while it amounted to $N$=$8.0 \times 10 ^{17} cm^{-3}$ for 6T 
using $t_{s}$= 5.8 nm at $V_{G}$=15 V.

On the other hand, the concentration of the mobile major carriers can be estimated by the 
conductance of TFT at $V_{G}$=0 V. Here we assume zero contact potential between organic 
semiconductors and Au electrodes. This assumption is supported by ohmic behavior of the S, D 
contacts to the pentacene and 6T active layers observed in the present FET characteristics. 
The conductivity of the 4.4 nm thick pentacene film was 
$ 2.5 \times 10 ^{-2} \Omega ^{-1} cm ^{-1}$ using $I_{SD}$ ($ 6.0 \times 10 ^{-7}$ A) at 
$V_{G}$=0 V, and the conductivity of the 5.8 nm thick 6T film was 
$ 1.6 \times 10 ^{-6} \Omega ^{-1} cm ^{-1}$  
using $I_{SD}$ ($5.1 \times 10 ^{-11}$ A) at $V_{G}$=0 V. 
Since the conductivity is represented as $ne \mu$, 
where $n$, $e$, and $\mu$ are carrier density, unit charge, and mobility, 
the concentration of the mobile major carriers ($N_{m}$) at $V_{G}$=0 V is obtained 
by dividing the conductivity by the mobility of the organic film. 
Therefore, $N_{m}$ was estimated to be 
$7.6 \times 10^{17} cm^{-3}$ for pentacene and $N_{m}$=$8.2 \times 10^{16} cm^{-3}$ for 6T. 

The fact that $N$ determined by $d_{th}$, and $N_{m}$ determined 
by the conductance at $V_{G}$=0 V, have the same order, 
indicates that $d_{th}$ actually corresponds to the depletion layer thickness. 
In other words, we could directly determine the deletion layer thickness 
as a shift of $d_{th}$ by the present {\it in-situ} and continuous measurement. 
Furthermore, the measurement was revealed to be a rather general method 
to determine the depletion layer thickness, 
because a shift of $d_{th}$ was observed for both organic semiconductors.

Although the above discussion on $d_{th}$ is roughly justified, 
there is a difference between $N$ and $N_{m}$; $N_{m}$ is always lower than $N$. 
The difference could be caused by several assumptions stated above 
(i.e. flat band condition, zero contact potential and depletion approximation). 
On the other hand, the difference could include important information 
on the concentration of the major carriers. 
Some of carriers are trapped at trap sites, such as grain boundaries, 
defects, or semiconductor/insulator (metal) interfaces. 
Therefore, the difference of concentration between $N$ and $N_{m}$ can be considered 
to relate to the concentration of the trapped carriers. 

Here we comment on the effect of impurity on the determination of the depletion layer 
thickness by the present measurement. 
Finite conductivity at $V_{G}$=0 V means that the carrier concentration 
originating from the impurity is higher than the concentration of trap sites. 
This causes positive threshold gate voltage($V_{th}$) in the FET characteristics. 
In case of purified samples, however, the carrier concentration 
can be lower than the concentration of trap sites, and $d_{th}$ would 
not be observed at positive $V_{G}$. 
High concentration of impurity is, thus, essential for the present method. 
Since the depletion layer thickness is inversely proportional to 
the concentration of major carriers, 
high concentration of impurity help determination of the depletion layer thickness 
($\sim$5 nm at $V_{G}$=15 V).

\subsection{Accumulation layer}
\label{sec3b}

In the previous section, we discussed the depletion layer thickness. 
In this section, we discuss the thickness of the accumulation layer, 
and the distribution function of charge injected by electric fields. 
In the following analysis, we assume that the organic semiconductors grow 
in a layer-by-layer fashion\cite{afm}, and that the carrier concentration depends 
only on electric fields ($V_{G}$), and does not depend on the film thickness. 
The observed conductance of TFT is the sum of the conductance of each layer, 
which is parallel to the interface. The conductance of each layer can be, 
thus, estimated by the first derivative of the total conductance 
with respect to film thickness. Then, the carrier density of each layer 
is determined by dividing the conductance of each layer by $\mu$. 
Based on the assumption of layer growth, the mobility can be constant, 
because mobility depends on morphology of the TFT\cite{9}. 
The mobility is determined by the FET characteristics of the 20 nm 
thick film (0.23 $cm^{2}V^{-1}s^{-1}$ for pentacene, 
$2.5 \times 10 ^{-4}cm^{2}V^{-1}s^{-1}$ for 6T). 

\begin{figure}
\begin{center}
\leavevmode\epsfysize=60mm \epsfbox{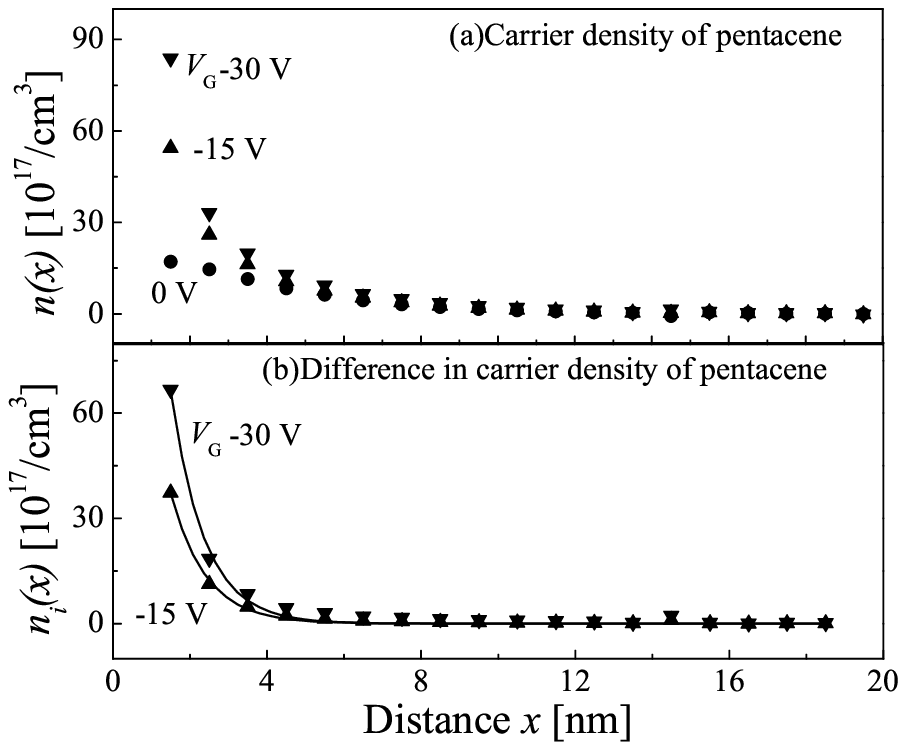}
\caption{(a) Carrier density ($n(x)$) of pentacene as a function of the distance ($x$) 
from the interface at various gate voltages ($V_{G}$); 
$V_{G}$=-30 V(down-triangle), -15 V(upper-triangle), 0 V(circle). 
(b) The difference ($n_{i}(x)$) between charge carrier density at $V_{G}$=-15 V or -30 V 
and that at $V_{G}$=0 V.}
\label{fig3}
\end{center}
\end{figure}

\begin{figure}
\begin{center}
\leavevmode\epsfysize=60mm \epsfbox{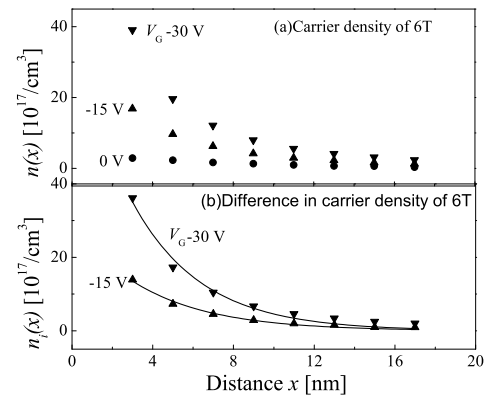}
\caption{ (a) Carrier density ($n(x)$) of 6T as a function of the distance ($x$) 
from the interface at various gate voltages ($V_{G}$); $V_{G}$=-30 V(down-triangle), 
-15 V(upper-triangle), 0 V(circle). (b) The difference 
($n_{i}(x)$)  between charge carrier density at $V_{G}$=-15 V or -30 V 
and that at $V_{G}$=0 V.}
\label{fig4}
\end{center}
\end{figure}

Figures~\ref{fig3} and ~\ref{fig4} show the obtained distribution function 
($n(x)$) of carriers as a function of distance $x$ from the interface 
for pentacene and 6T at various $V_{G}$. 
The large carrier density at $V_{G}$=0 V in small $x$ region is 
due to the charge transfer from the Au electrodes to the pentacene molecules \cite{10}. 
By subtracting the carrier density $n_{0}(x)$ at $V_{G}$=0 V from 
$n_{-15}(x)$ at $V_{G}$=-15 V, or $n_{-30}(x)$ at $V_{G}$=-30 V, 
the carrier density injected by electric fields ($n_{i}(x)$) is obtained. 
As seen in the figure, $n_{i}(x)$ decays steeply with increasing $x$, 
meaning that the injected carriers are localized at the interface. 
We could quantitatively determine the distribution function of carriers and the 
accumulation layer thickness for organic TFT by the {\it in-situ} and continuous measurement. 
It should be noticed that the above processes (differentiation, division by $\mu$) 
can be validated only for the continuous measurement results, 
in which the sample condition is kept unchanged. 
Otherwise, for an individual sample, the condition changes from one sample 
to another due to the difference in grain size, crystallinity, interface state, etc.

For quantitative estimation on the accumulation layer thickness, $n_{i}(x)$ is fitted with an 
exponential function ($f(x)=a \times exp(-x/b)$). 
The fitted value of $b$ was 2($\pm$1) nm for pentacene and 9($\pm$5) nm for 6T 
at both $V_{G}$=-15 V and -30 V \cite{11}. 
Again, the value of $b$ was well reproduced for several experiments.
Here, $b$ can be considered as an effective accumulation layer thickness. 
There are three points to note about the accumulation layer thickness ($d_{ac}$). 
First, most of injected carriers are localized in a few ML next to the interface.
(1ML thickness: 1.5 nm for pentacene, 2.2 nm for 6T)
We could quantitatively show the localization of injected carrier at the interface by 
experiments, while most previous works simply shows 
that carriers were localized at the interface\cite{5}. 
Second, $d_{ac}$ does not depend on gate voltages. 
Third, $d_{ac}$ of pentacene is smaller than that of 6T. 
In the following, we would discuss these gate voltage and chemical species dependence of the 
accumulation layer thickness.

First, we discuss the gate voltage dependence of the accumulation layer thickness. 
In an ideal semiconductor, $d_{ac}$ is inversely proportional to $V_{G}$\cite{12}. 
On the other hand, $d_{ac}$ of organic semiconductor pentacene and 
6T did not show gate voltage dependence in the present study. 
The difference could be explained by the electron-electron repulsion, 
the localized states in organic semiconductors, and geometry-related factors. 
Electron-electron repulsion delocalizes the carriers, and leads to an increase in $d_{ac}$. 
Since the repulsion force increases with carrier concentration (i.e. $V_{G}$), 
a decrease of $d_{ac}$ could be compensated with an increasing $V_{G}$. 
This causes little $V_{G}$ dependence. 
On the other hand, $V_{G}$ dependence in an ideal semiconductor assumes 
the parabolic density of states for the conduction band. 
In organic TFT, however, there is no long range order in the specimens. 
Because of absence of the long range order, 
the localized states with finite energy width are formed below the free 
electron like conduction band. Therefore, the equations assuming parabolic DOS could not be 
applicable to the organic TFT, and $d_{ac}$ was not inversely proportional to $V_{G}$.    
The geometry-related factors can also explain the experimental results. 
There are many grain boundaries in the growth film, 
and these grain boundaries and defects would affect the charge 
injection. Therefore, the equations assuming a perfect semiconductor crystal 
could not be applicable to the organic FET. 
Although we present some explanation of the little gate voltage dependence of 
the accumulation layer thickness, we can not present decisive one at present.

Second, the chemical species dependence of the accumulation thickness is discussed. 
For p-type organic semiconductors, holes are injected 
in the highest occupied molecular orbital (HOMO) of the semiconductor, 
and thus, $d_{ac}$ reflects the spatial distribution of HOMO. 
On the other hand, both pentacene and 6T are $\pi$-conjugated molecules, 
and their HOMOs spread in the whole molecules. 
Therefore, the molecular size should be taken into account to discuss $d_{ac}$. Since the 
molecular size of 6T is larger than that of pentacene, 
$d_{ac}$ of 6T was larger than that of pentacene. 
Besides the molecular size, another factors, such as difference in dielectric constant, 
intrinsic carrier density, could cause the difference in $d_{ac}$\cite{13}.

Here we estimate the amount of carriers injected by electric fields at the first layer, 
in order to obtain a guide to the electric fields control of physical properties. 
Since the distribution function of the injected carrier can be represented by 
$f(x)$=$a \times exp(-x/b)$, the amount of carriers in the first 
layer is obtained by integrating $f(x)$ from 0 to the thickness of 1 ML 
(1.5 nm for pentacene, 2.2 nm for 6T ). 
The amount of the injected carriers was $6.1 \times 10^{11}cm^{-2}$ for pentacene 
and $2.6 \times 10^{11}cm^{-2}$ for 6T at $V_{G}$=-30 V. 
One carrier was injected every 350 (810) pentacene (6T) molecules at $V_{G}$=-30 V.  
In most of previous studies, the amount of injected carriers per each molecule was estimated 
assuming that carrier were localized at the first layer. 
In the present work. however, distribution of injected carriers were estimated, 
showing that they were not necessarily localized at the first layer. 
The present result can be, thus, 
a guide to control of physical properties by electric fields. 

In the present study, the {\it in-situ} and continuous measurement was applied to organic 
semiconductor, in order to quantitatively determine the thickness 
of the depletion and accumulation layers. 
As discussed in the previous section, layer growth is essential to the present method. 
The reason why we studied organic semiconductors, 
is just that they do not need high substrate temperature to obtain flat films, 
in contrast with the inorganic semiconductor which needs high substrate temperature. 
Only if we can prepare flat films at low substrate temperature, the present 
method can be applied to inorganic semiconductor to estimate 
the thickness of the depletion and accumulation layers, 
and distribution function of the injected carriers.

Finally, we comment the control of physical properties by electric fields. 
Since carriers reside in a few ML from the semiconductor/insulator interface, 
the physical properties only at the interface could be controlled by the charge injection. 
This implies the importance of a well-ordered interface 
through which charges are efficiently injected. 
Furthermore, we propose that the semiconductor-metal or -superconductor transition 
is a promising target to study. When the 
semiconductors change into metal phase by field induced charge injection, 
metallic and semiconducting regions sit side by side 
with only an atomic distance apart in the organic film. 
Under this situation, free carriers can interact with exciton, 
associated with the semiconductor, at the interface. 
Ginzburg {\it et al.}\cite{14} proposed the possibility of superconductivity by exciton 
mechanism at metal/semiconductor interfaces. 
They discussed that the coexistence (in real space) of 
excitons and metallic carriers would make it possible interface superconductivity. 
Therefore, the situation realized at interface thus leads a possible ground 
for superconductivity by exciton mechanism.

\section{CONCLUSIONS}
\label{sec4}
We present the {\it in-situ} and continuous measurement of the conductivity 
of growing organic films, as a simple but powerful method to determine 
the distribution curve of injected carriers and the 
dimension of the accumulation and depletion layers. 
The depletion layer thickness could be directly 
determined as a shift of the threshold thickness at which electric current began to flow. 
The {\it in-situ} and continuous measurement 
could also determine qualitatively the accumulation layer thickness 
together with the distribution function of injected carriers. 
The accumulation layer thcikness was 2 nm for pentacene and 9 nm for 6T.
The accumulation layer thickness did not depend on gate voltages, 
rather depended on the chemical species.

\acknowledgments{This work was supported by the Grant-in-Aid 
from the Ministry of Education, Culture, Sports, 
Science and Technology of Japan (14GS0207).}

\end{multicols}

\begin{references}
\bibitem{4}The depletion layer thickness can be indirectly determined by the capacitance measurement. 
When the depletion layer is formed at the semiconductor/insulator interface, the capacitance of the system 
is $ C_{i}C_{d}/(C_{i}+C_{d})$, where $C_{i}$ and $C_{d}$ are the capacitance of the insulating and deletion layers. 
The capacitance of the depletion layer is determined using the known capacitance of the insulating layer. 
Therefore, the depletion layer thickness ($L_{d}$) can be determined by equation $L_{d}=\epsilon/C_{i}$, 
where $\epsilon$ is the dielectric constant of the semiconductor. 
On the other hand, the accumulation layer thickness is hard to be determined. 
In most of previous studies of organic FET, it is assumed that the whole charge (=$C_{i}V_{G}$) 
is injected only in the first layer at the interface.
\bibitem{afm} This assumption is supported by the AFM results. 
The film nearly covered the substrate completely, although there were some grain boundaries. 
Furthermore the deviation of height in the island was about 2 nm, 
which corresponded to 1 monolayer thickness. These experimental results supports 
that the nearly ideal layer growth occurred under the present growth condition.
\bibitem{11}The carrier density ($n(x)$) could be also obtained 
by dividing the local conductance by the mobility $\mu _{f}(x)$, 
which is determined from the FET characteristics at each film thickness. 
In this case, $b$ is 0.86 nm at -15 V, and 0.91 nm at -30 V for pentacene, 
$b$ is 3.4 nm at -15 V, and 3.9 nm at -30 V for 6T, 
which are smaller than $b$ in the text. In determining $\mu _{f}$, 
the carriers($Q=CV$) are assumed to be completely injected 
even for an ultra thin film. The injected carriers are, thus, 
overestimated, leading to underestimation of $\mu _{f}$. 
On the other hand, constant mobility ($\mu _{c}$) in the text overestimates 
the mobility especially in the thin film region. Therefore, the true mobility is 
between $\mu _{f}$ and $\mu _{c}$, and the true $b$ is between those estimated 
from $\mu _{f}$ and $\mu _{c}$. 
The main topic of the present work is based on the fact that $b$ does not depend on $V_{G}$, 
which was commonly observed in both cases. 
Therefore, the discussion in the text was not affected by choice of $\mu$.
\bibitem{13}The carrier density is represented as  $p=n_{i}exp((E_{i}-E_{F})/kT)$, 
where $n_{i}$, $E_{i}$, $E_{F}$ are intrinsic carrier density, 
intrinsic Fermi level, and Fermi level. Since the carrier density changes exponentially 
with the energy difference between intrinsic Fermi level and Fermi level, 
the small difference in the band bending would lead to large difference 
in distribution function of carriers. The band 
bending depends on dielectric constant of semiconductors.
\end{references}
\end{document}